\documentclass[a4paper,superscriptaddress,prd,showpacs,preprintnumbers,10pt,twocolumn,eqsecnum,reprint]{revtex4}
\pdfoutput=1
\usepackage{amsmath,color}
\usepackage{color}
\usepackage{amsmath,amsfonts,amssymb,mathrsfs}
\usepackage{booktabs}
\usepackage{graphicx}
\usepackage{float} 	
\usepackage{array}
\usepackage{longtable}
\allowdisplaybreaks
\AtBeginDocument{
\heavyrulewidth=.08em
\lightrulewidth=.05em
\cmidrulewidth=.03em
\belowrulesep=.65ex
\belowbottomsep=0pt
\aboverulesep=.4ex
\abovetopsep=0pt
\cmidrulesep=\doublerulesep
\cmidrulekern=.5em
\defaultaddspace=.5em
}

\newcommand\this{\addtocounter{equation}{1}\tag{\theequation}}

\usepackage{hyperref}								
\hypersetup{bookmarksopen=false,colorlinks,pdfstartview=FitH, citecolor=black,runcolor=black,anchorcolor=black,linkcolor=black,
			pdftitle = {PaperTitle}
   		 	}
   		 	
\begin{document}

\title{Non-Abelian quantum field theory of pionic strong interactions}

\author{C. A. Dominguez}
\affiliation{Centre for Theoretical \& Mathematical Physics, and Department of Physics, University of
Cape Town, Rondebosch 7700, South Africa.}%

\author{P. Moodley}
\affiliation{Centre for Theoretical \& Mathematical Physics, and Department of Physics, University of
Cape Town, Rondebosch 7700, South Africa.}

\author{K. Schilcher}
\affiliation{Centre for Theoretical \& Mathematical Physics, and Department of Physics, University of
Cape Town, Rondebosch 7700, South Africa.}
\affiliation{PRISMA Cluster of Excellence, Institut f\"{u}r Physik,\\ Johannes Gutenberg-Universit\"{a}t
Staudingerweg 7, D-55099 Mainz, Germany.}
\affiliation{National Insitute for Theoretical Physics, Private Bag X1, Matieland 07602, South Africa.}

\author{G. B. Tupper}
\affiliation{Centre for Theoretical \& Mathematical Physics, and Department of Physics, University of
Cape Town, Rondebosch 7700, South Africa.}

\date{\today}

\begin{abstract}
A renormalizable non-Abelian theory of strong interactions of pions, mediated by rho-mesons, is
formulated at tree- and at one-loop level in perturbation theory. Hadron masses are generated through
spontaneous symmetry breaking using the Higgs mechanism. Quantization and gauge fixing is
achieved using the generalized class of $R_\xi$ gauges. As an application of this theory,  pion-pion
scattering lengths are obtained at tree-level in good agreement with data.
\end{abstract}

\maketitle

\section{Introduction}
A viable Abelian renormalizable quantum field theory of pionic strong interactions mediated by a neutral, massive rho-meson
was proposed long ago by Kroll, Lee and Zumino (KLZ) \cite{KLZ1967,Lowenstein1972}. The Lagrangian has the form
\begin{align*}
\mathscr{L} &= {\partial _\mu }\phi\, {\partial ^\mu }{\phi ^*} - {m^2}\phi \, {\phi ^*} - \frac{1}{4}{F_{\mu \nu }}\,{F^{\mu \nu }} + \frac{1}{2}{M^2}\,{A_\mu }\,{A^\mu }  \\
 &+ i\, {g_{\rho \pi \pi }} \,{A_\mu }\, J_\pi ^\mu  + g_{\rho \pi \pi }^2\, {A_\mu }\,{A^\mu }\phi \,
 {\phi ^*} \,.\this
\end{align*}
with $m$ and $M$ the pion and rho-meson masses, respectively, $A_\mu$ a vector field of the $\rho_0$ meson $(\partial_\mu A^\mu = 0)$, $\phi$ a complex pseudo-scalar field describing the $\pi^\pm$ mesons, $F_{\mu\nu}$  the field strength tensor,  $F_{\mu\nu} =\partial_\mu A_\nu-\partial_\nu A_\mu$, and
$J_\pi ^\mu $ the $\pi^\pm$ current, $J_\pi ^\mu  = {\phi ^*}  \overleftrightarrow{\partial ^\mu }  \phi $,  with ${\partial _\mu }J_\pi ^\mu  = 0$.
Renormalizability is ensured by the massive gauge boson being coupled to a conserved current \cite{vanHees2003,RUEGG2004}. This 
theory provides the necessary platform supporting the purely phenomenological Vector Meson Dominance Model
(VMD) \cite{SAKURAI1960,Sakurai1969}. Examples of successful applications of the KLZ theory are the electromagnetic form factor 
of the pion in the time-like region \cite{GALE1991}, as well as in the space-like region \cite{CADWillers2007}, and the pion 
scalar radius \cite{CADWillers2008} and scalar form factor in the space-like region \cite{CADLushozi2015}.

The Abelian nature of this theory limits the range of applications, and precludes its candidacy for becoming
a viable theory covering the intermediate energy region between threshold, dominated by chiral perturbation theory (CHPT), and the onset of perturbative QCD.  A
clear way forward is to formulate a non-Abelian, renormalizable quantum field theory of pionic interactions, invoking 
spontaneous symmetry breaking to generate the mass of the gauge bosons. This approach is described in some detail here, at the tree{\color{red}-}level in perturbation theory. As an application of this theory, the pion-pion scattering lengths are 
obtained, at tree-level, in good agreement with data. A one-loop determination would involve considerable work, well beyond the 
scope of this paper

\section{SU(2) Extension}
A generalization of the Abelian KLZ theory must accommodate the full triplet of 
pseudoscalar pions $\{\pi^-,\pi^0,\pi^+ \}$, and vector rho-mesons $\{\rho^-,\rho^0,\rho^+\}$. To accomplish this generalization  some approximations to simplify the process are required. For instance, one can ignore the small mass differences between the charged and uncharged particles. These are $m_{\pi^\pm} - m_{\pi^0} = 4.5936 \pm 0.0005  ~\rm{MeV}$, and  $m_{\rho^\pm} - m_{\rho^0}= 0.7\pm 0.8 ~ \rm{MeV}$, i.e. negligible on a hadronic scale. This degeneracy in masses allows for the use of the larger Isospin gauge group  SU(2). The gauge principle  guides  in the construction of a Lagrangian which is invariant under SU(2) gauge transformations, and is renormalizable. This naturally leads to a Yang-Mills type Lagrangian. The pion field is labelled as $\Phi(x)\equiv\phi^a \in \{\phi^1,\phi^2,\phi^3\} $ and the rho-meson field as $A_\mu(x) \equiv A^a_\mu \,T^a$, with $A^a_\mu \in \{A^1_\mu,A^2_\mu,A^3_\mu\}$.  The gauge transformation is
\begin{align}
	U&=e^{i\alpha^a(x) T^a} \in \; \rm{SU(2)} \,,\\
	\Phi &\to \Phi' = U\Phi, \\
	A{_\mu } &\to A{'_\mu } = UA_\mu{U^\dag } + \frac{1}{{i \,e}}U{\partial _\mu }{U^\dag } \,,
\label{gaugetrans}
\end{align}
with the covariant derivative being
\begin{align}
{D_\mu } = {\partial _\mu } + i\, e\, {T^a}A_\mu ^a \,,
\end{align}
and $a \in\{1,2,3\}$. Here $e\equiv g_{\rho\pi\pi}$ is the dimensionless gauge coupling, and $T^a$ belongs to the SU(2) Lie algebra. We define
\begin{align}
	F_{\mu \nu }^a \equiv {\partial _\mu }A_\nu ^a - {\partial _\nu }A_\mu ^a
	\,,
\end{align}
so that the non-Abelian field strength tensor is
\begin{align}
	G_{\mu \nu }^a = F_{\mu \nu }^a - e \,{\varepsilon _{abc}} \,A_\mu^b \, A_\nu ^c \,.
\end{align}
The locally SU(2) gauge invariant Lagrangian becomes
\begin{align}
 \mathscr{L} &= \frac{1}{2}
 {\left( {{D_\mu }\Phi } \right)^\dag }{D^\mu }\Phi    - \frac{1}{4}{G^a}_{\mu \nu }{G^a}^{\mu \nu }\,.
 \label{flag}
\end{align}
The gauge principle also allows for the inclusion of a polynomial with an infinite number of terms of the form
\begin{align}
P(\Phi)&=\sum\limits_{n=1}^{\infty}c_n(\Phi^\dag \Phi)^n \nonumber \\
&= - \frac{1}{2} b^2\Phi^\dag \Phi+ \frac{\lambda_4}{8}(\Phi^\dag \Phi)^2+\sum\limits_{n=3}^{\infty}c_n(\Phi^\dag \Phi)^n \,.
\label{polynomial}
\end{align}
This pionic quartic coupling term  was left out in the  U(1) KLZ model \cite{KLZ1967}.
There is no legitimate reason at this stage to exclude this term, so it will be kept in the Lagrangian. For terms with $n>3$, the $c_n$ have dimension of inverse mass, ${\mbox{Dim}}\,[c_n]=M^{-n}$. This poses a problem for the requirements of a renormalizable theory \cite{Bohmbook}-\cite{BardinbookSM}. Hence, terms with $n\ge 3$ will not be considered. The Lagrangian then takes the form
\begin{align*}
\mathscr{L}_{\phi A}  &= \frac{1}{2}\left( {{\partial _\mu }{\phi _a}{\partial ^\mu }{\phi _a} + {b^2}\phi _a^2} \right) - \frac{1}{2}{e^2}{\varepsilon _{aij}}{\varepsilon _{bjk}}{\phi _i}{\phi _k}A_\mu ^a{A^{b\mu }}
\\
&+ \frac{1}{2}e{\varepsilon _{aij}}A_\mu ^a\left( {{\phi _j}{\partial ^\mu }{\phi _i} - {\phi _i}{\partial ^\mu }{\phi _j}} \right)  - \frac{{{\lambda _4}}}{8}\left( {\phi _a^2\phi _b^2} \right)  \\
 &- \frac{1}{4}F_{\mu \nu }^a{F^{a\mu \nu }} + \frac{1}{2}e{\varepsilon _{abc}}{A^{b\mu }}{A^{c\nu }}F_{\mu \nu }^a \\
 &- \frac{1}{4}{e^2}{\varepsilon _{abc}}{\varepsilon _{ade}}A_\mu ^bA_\nu ^c{A^{d\mu }}{A^{e\nu }} \this .
 \label{mlesslag}
\end{align*}

\section{Spontaneous Symmetry breaking}
The rho-meson mass will be generated dynamically through spontaneous symmetry breaking, invoking the Higgs mechanism. The Lagrangian, Eq.\eqref{mlesslag} would accomplish this but it would break Isospin symmetry.
 Thus, we introduce a complex doublet of fields $X$, which satisfies the gauge transformation
\begin{align}
X  \to X ' = UX \,,
\end{align}
and construct an SU(2) gauge invariant Higgs Lagrangian,
\begin{align}
\mathscr{L} = \left(D_\mu X\right)^\dag \left(D^\mu X\right) - V\left(X,X^\dag\right) - \kappa \left(X^\dagger X \right) \left(\Phi^\dagger\Phi\right)\,,
\label{higgspotbsb}
\end{align}
where the potential $V\left(X,X^\dag\right)$ is 
\begin{align}
V\left(X,X^\dag\right) = \dfrac{\lambda}{8} \left(X^\dagger X \right)^2 - \dfrac{\mu^2}{2}\left(X^\dagger X \right) \,,
\end{align}
which has been tuned for symmetry breaking with $\lambda>0$, and $\mu^2>0$. The term $\kappa \left(X^\dagger X \right) \left(\Phi^\dagger\Phi\right)$ is present as it is gauge invariant, with $\kappa$ a dimensionless coupling. The doublet is parametrized as
\begin{align}
X = \dfrac{1}{\sqrt{2}} \begin{pmatrix}
										x_2 + i x_1 \\
										x_0 - i x_3
							\end{pmatrix} \,.
\end{align}
Defining $\nu^2 \equiv \dfrac{\mu^2}{\lambda}$, the $X$ field  acquires its vacuum expectation value with the above choice of a minimum. The vacuum configuration is
\begin{align}
X_\text{M} =  \dfrac{1}{\sqrt{2}} \begin{pmatrix}
0 \\
2\nu
\end{pmatrix} \,.
\label{vevlocation}
\end{align}

Considering fluctuations around the minimum $X_\text{M}$ we define
\begin{align*}
 \chi \equiv X - X_\text{M} \this \,,
 \label{Xchidef}
\end{align*} 
where $\chi$ is the new field perturbed around the vacuum, and parametrized as
\begin{align*}
\chi =  \dfrac{1}{\sqrt{2}} \begin{pmatrix}
\chi_2 + i \chi_1 \\
\chi_0 - i \chi_3
\end{pmatrix} \,, \label{chitrans}\this
\end{align*}
Expressing the Lagrangian, Eq.\eqref{higgspotbsb}, in terms of this new translated field and relabelling  $\chi_0$ as ${\chi _0} \equiv H$, which we refer to as the Higgs field, leads to
\begin{align*}
&\mathscr{L}_H = \frac{1}{2}\left( {{\partial _\mu }H{\partial ^\mu }H - {\mu ^2}{H^2}} \right) + \frac{1}{2}{\partial _\mu }{\chi _a}{\partial ^\mu }{\chi _a} \\
	&+ ie{\partial ^\mu }A_\mu ^a\left[ {X_{\rm{M}}^\dag {T^a}\chi  - {\chi ^\dag }{T^a}{X_{\rm{M}}}} \right]  -\frac{1}{4}e{\varepsilon _{abc}}\left( {{\chi _a}\overleftrightarrow{\partial^\mu}{\chi _b}} \right)A_\mu ^c \\
&- \frac{1}{2}e\left( {{\chi _a}\overleftrightarrow{\partial^\mu}H} \right)A_\mu ^a +
 \frac{1}{2}{e^2}{\nu ^2}A_\mu ^a{A^{a\mu }} + \frac{1}{2}{e^2}\nu HA_\mu ^a{A^{a\mu }} \\ 
 &+ \frac{1}{8}{e^2}{H^2}A_\mu ^a{A^{a\mu }}  + \frac{1}{8}{e^2}\chi _b^2A_\mu ^a{A^{a\mu }} - \frac{\lambda }{{32}}{H^4} - \frac{\lambda }{{16}}{H^2}\chi _a^2  \\
&- \frac{\lambda }{{32}}\chi _a^2\chi _b^2 - \frac{{\nu \lambda }}{4}{H^3} - \frac{{\nu \lambda }}{4}H\chi _a^2 - \frac{1}{2}\left( {4\kappa {\nu ^2}} \right){\phi^2_a }  \\ &- 2\kappa \nu H{\phi^2_a }  - \frac{1}{2}\kappa {H^2}{\phi^2_a }  - \frac{1}{2}\kappa \chi _a^2{\phi^2_a }  + \frac{1}{2}{\mu ^2}{\nu ^2} \,.\this 
\label{higgslagrange}
\end{align*}

Considering small perturbations around the vacuum field configuration generates the mass terms for the vector field, $\frac{1}{2}{e^2}{\nu ^2}A_\mu ^a{A^{a\mu }}$. A contribution to the potential of the pion was generated during the symmetry breaking process, $\frac{1}{2}\left( {4\kappa {\nu ^2}} \right){\phi^2_a }$. A mass term for the Higgs field was also generated in this process, $\frac{1}{2}{\mu ^2}{H^2}$. The $\chi_a$ are the three massless Goldstone fields. The classical Lagrangian is then given as the sum of the pion-rho Lagrangian and the Higgs Lagrangian 
\begin{align}
\mathscr{L}_{\rm cl} \equiv \mathscr{L}_{\phi A} + \mathscr{L}_H
\label{claslag}
\end{align}
%
\section{Faddeev-Popov Ghosts and Gauge Fixing}
The classical Lagrangian, Eq.(\ref{claslag}) is now ready for quantization. The gauge transformation partitions the configuration space of fields. Thus it sets up an equivalence class for sets of physically equivalent fields. Since there are infinitely many field configurations related to each other via a gauge transformation, this leads to a divergence when summing over all field contributions. Hence one needs to count only one member from each partition. This is done through a gauge fixing function designed to span the configuration space, and intersect the set of all physically equivalent fields only once.
This is implemented using the identity
\begin{align}
\Delta \left[ A \right]\int {\mathcal{D}\mu \,[U] \,\delta \left[ {{G^a}\left( {{A[U]}} \right) - {w^a}} \right]}  = 1 \,,
\label{fpid}
\end{align}
where $\Delta \left[ A \right]$ is a determinant, and $\mathcal{D}\mu [U]$ is the invariant Haar measure \cite{Bohmbook}.  Inserting Eq.\eqref{fpid} into the path integral, and exploiting its invariance under gauge transformations, allows for the extraction of the multiplicative divergence.
Dividing out the Haar volume, one can define a path integral not affected by over-counting as
\begin{align}
\mathcal{Z} &= \int {\mathcal{D}A \, e^{iS\left[ A \right]}}
\exp \left[ - \frac{i}{{2\xi }}\,\int {{d^4}x} \,G_a^2\,\left(A\right)\right] \Delta \left[A\right]  \;, 
  \label{pathintegral}
\end{align}
where $\Delta[A]$ is the Faddeev-Popov determinant \cite{FADDEEV1967}
\begin{align*}
\Delta \left[ A \right] &= \det {\left[ {\frac{{\delta {G^b}\left[ {^\alpha A;x} \right]}}{{\delta {\alpha ^c}\left( y \right)}}} \right]} 
 = \det {M^{bc}}\left( {x,y} \right) \,. \this
\end{align*}
One can choose a class of gauge fixing functions of the form \cite{FujikawaLee,Abers19731}
\begin{align}
{G^b}\left[ {^\alpha A;x} \right] = {\partial ^\mu }A_\mu ^b + \xi ie\left( {X_{\rm{M}}^\dag {T^b}\chi - {\chi^\dag }{T^b}X_{\rm{M}}} \right)\,. 
\label{gffunction}
\end{align}
Computing the variational derivative of the gauge fixing function with respect to the group parameters gives
\begin{eqnarray}
{M^{bc}} &=& \left[ - \partial _x^\mu \, D_\mu ^{bc} - \xi\, {e^2}\,{\nu ^2} \,{\delta ^{bc}} - \frac{1}{2}\,\xi\, {e^2}\,\nu\, {\chi _0}\,{\delta ^{bc}} 
\nonumber \right.\\ [.3cm]
&+& \left. \frac{1}{2}\,\xi {e^2}\,\nu \,{\chi _a}\,{\varepsilon _{abc}} \right] \,  \frac{1}{e}\delta \left( {x - y} \right) \,.
\end{eqnarray}
The Faddeev-Popov determinant can be expressed as a Berezin-type functional integral \cite{BerezinbookSuperAnalysis} over Grassmann fields
\begin{align*}
&\Delta \left[ A \right] =\det \left[ {{M^{bc}}} \right]\\
 &= \int {\mathcal{D}\bar u \, \mathcal{D}u \, \exp\left[ - i e\int {{d^4}x \,{d^4}y \,{{\bar u}^b}\left(x\right){M^{bc}}\left({x,y} \right)\,{u^c}\left(y\right)}\right]} \this\,,
\end{align*}
where $\bar{u}^a, {u}^a$ are the anti-commuting Grassmann fields (Faddeev-Popov ghosts). Substituting the determinant into the path integral leads to
\begin{align*}
\mathcal{Z} 
 &= \int {\mathcal{D}\,\bar u\mathcal{D}u \,\mathcal{D}A} \; \exp\, \left( i\int {{d^4}x \,{\mathscr{L}_{{\rm{eff}}}}} \right) \,, \this
\end{align*}
where 
\begin{align*}
{\mathscr{L}_{{\rm{eff}}}} & = \mathscr{L_{\rm{cl}}} - \frac{1}{{2\xi }}G_a^2\left( A \right) + {{\bar u}^a}\left( {{\partial ^\mu }{\partial _\mu } + \xi \, {e^2} \, {\nu ^2}} \right){u^a} \\ 
&+ \,e \,{\varepsilon _{abc}} \, A_\mu ^a\left( {{\partial ^\mu }{{\bar u}^b}} \right){u^c} + \frac{1}{2}\,\xi \,{e^2} \,\nu\, {\chi _0}{{\bar u}^a}{u^a}\\ 
&- \frac{1}{2} \,\xi \,{e^2}\nu \,{\varepsilon _{abc}} \,{\chi _a}{{\bar u}^b}{u^c} \label{leff} \,.\this
\end{align*}

\section{Scattering Lengths}
We consider the elastic  scattering process 
\begin{equation}
\phi^a\left(p_1\right) + \phi^b\left(p_2\right) \to \phi^c\left(p_3\right) + \phi^d\left(p_4\right)\,, \label{Eq,5.1}
\end{equation}
where particles are external asymptotic states, i.e. $p_i^2=m^2$. We label the incoming and outgoing three momentum in the centre of mass frame as $\bf q$ and $\bf q'$, respectively, with the constraints
\begin{align}
  \bf{q}  &\ne \bf{q'} \quad\text{and} \quad
 |\bf q| = |\bf q '| \,.
\end{align}
Next, the cosine of the scattering angle and the ratio $R$ are
\begin{align}
z &= \cos \theta	\,,
&R = \frac{\bf|q|^2}{m^2} \label{zcostheta} \,.
\end{align}
The Mandelstam invariants in terms of $z$ and $R$ are
\begin{align}
 s &= 4{m^2}\left(1+R\right) \label{mandletams}\,,\\ 
 t &=  - 2{m^2}\left( {1 - z} \right)R \label{mandletamt}\,,\\ 
 u &=  - 2{m^2}\left( {1 + z} \right)R \label{mandletamu} \,.  
\end{align}

\subsection{Tree-Level Scattering Lengths}
\begin{figure}[htb]
\centering
\includegraphics[width=1.0\linewidth]{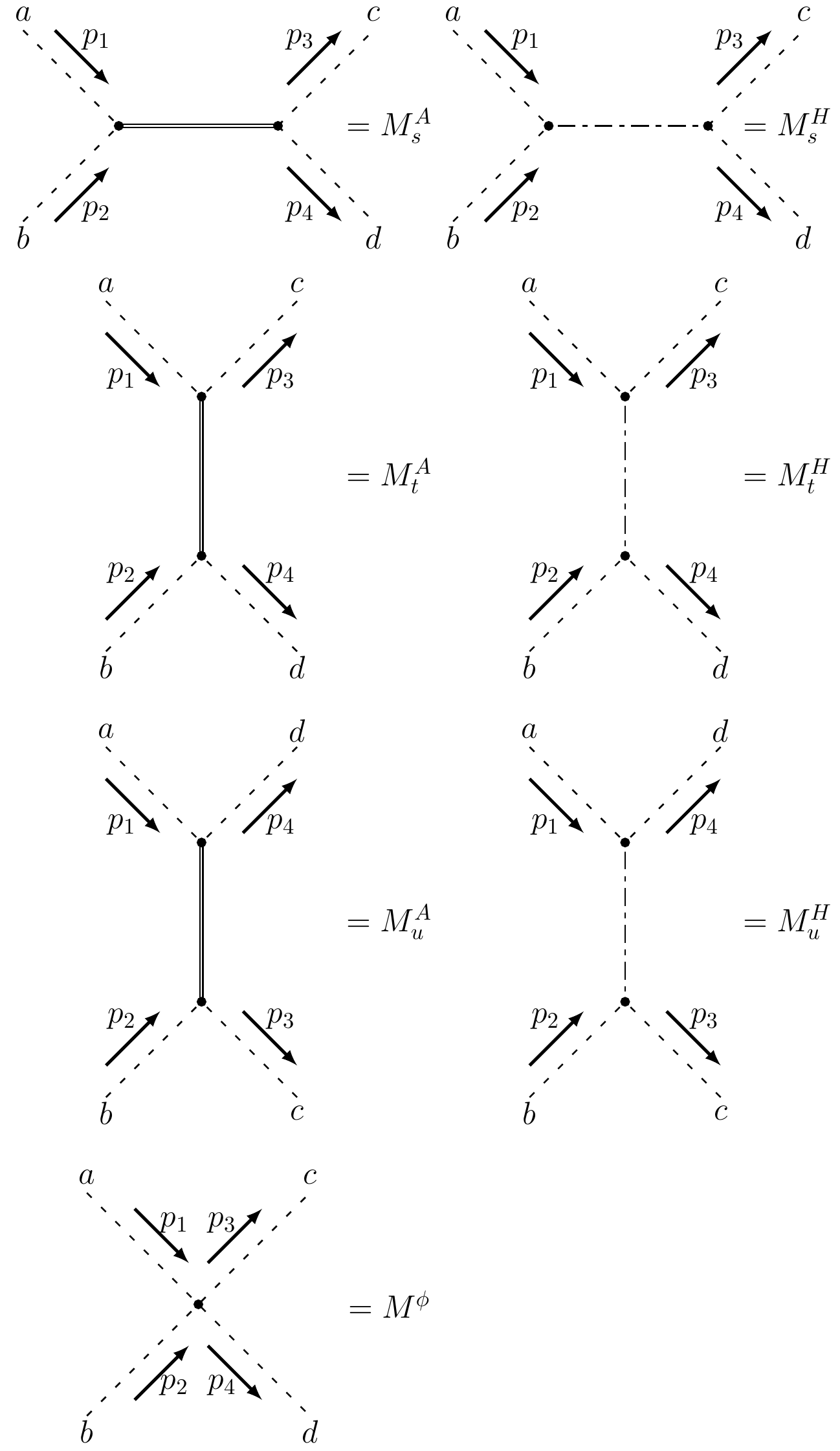}
\caption{Tree-level $\pi-\pi$ scattering processes.}
\label{fig AllTreeScatProc}
\end{figure}

We are now in a position to calculate the pion-pion scattering lengths. The  required tree-level diagrams generated from the Feynman rules are shown in figure \ref{fig AllTreeScatProc}. 
The $s$ channel scattering process mediated by the rho meson leads to the amplitude
\begin{align*}
M_s^A &= {S_A}\left( {{\delta _{ad}}{\delta _{bc}} - {\delta _{ac}}{\delta _{bd}}} \right) \,,\this \\
{S_A}&=  - 4\,i\,{e^2}\,{m^2} \,\frac{{z\, R}}{{4\,{m^2}\left( {1 + R} \right) - {M^2}}} \,.\this
\end{align*}
In the  $t$ channel it becomes
\begin{align*}
M_t^A &= {T_A}\left( {{\delta _{ab}}{\delta _{cd}} - {\delta _{bc}}{\delta _{ad}}} \right) \,,\this \\
{T_A} &= 2 \,i\,{e^2}\,{m^2} \,\frac{{2 + \left( {3 + z} \right)R}}{{2\,{m^2}\left( {1 - z} \right)R + {M^2}}} \,,\this
\end{align*}
and in the $u$ channel 
\begin{align*}
M_u^A &= {U_A}\left( {{\delta _{ab}}{\delta _{cd}} - {\delta _{ac}}{\delta _{bd}}} \right) \,,\this \\
{U_A}&= 2\,i\,{e^2}\,{m^2}\frac{{2 + \left( {3 - z} \right)R}}{{2\,{m^2}\left( {1 + z} \right)R + {M^2}}} \,.\this
\end{align*}
The total amplitude is then
\begin{align*}
T_A^{ab,cd} &= \left( {{T_A} + {U_A}} \right){\delta _{ab}}{\delta _{cd}} - \left( {{S_A} + {U_A}} \right){\delta _{ac}}{\delta _{bd}} \\
&+ \left( {{S_A} - {T_A}} \right){\delta _{ad}}{\delta _{bc}} \,.\this
\end{align*}
For the $s, t$ and $u$ channel scattering processes mediated by the Higgs-boson the results are
\begin{align}
M_s^H &= {S_H}{\delta _{ab}}{\delta _{cd}} \,, & {S_H}&=  - i\frac{{16{\kappa ^2}{\nu ^2}}}{{s - m_H^2}} \,,
\end{align}
\begin{align}
M_t^H  &= {T_H}{\delta _{ac}}{\delta _{bd}}\,, & {T_H}&=  - i\frac{{16{\kappa ^2}{\nu ^2}}}{{t - m_H^2}} \,,
\end{align}
\begin{align}
M_u^H  &= {U_H}{\delta _{ad}}{\delta _{bc}} \,, & {U_H}&=  - i\frac{{16{\kappa ^2}{\nu ^2}}}{{u - m_H^2}} \,.
\end{align}
The total amplitude due to the Higgs-boson is 
\begin{align}
T_H^{ab,cd} = {S_H}{\delta _{ab}}{\delta _{cd}} + {T_H}{\delta _{ac}}{\delta _{bd}} + {U_H}{\delta _{ad}}{\delta _{bc}} \,.
\end{align}
Finally, the amplitude due to  four-pion scattering is
\begin{align}
T_\lambda^{ab,cd} &={S_\lambda }\left( {{\delta _{ab}}{\delta _{cd}} + {\delta _{ad}}{\delta _{bc}} + {\delta _{ac}}{\delta _{bd}}} \right), &{S_\lambda }& =  - i{\lambda _4} \,.
\end{align}

\subsection{Isospin Amplitudes}
The most general form of the scattering amplitude is
\begin{align*}
{M^{ab,cd}} &= F\left( {s,t,u} \right){\delta _{ab}}{\delta _{cd}} + G\left( {s,t,u} \right){\delta _{ac}}{\delta _{bd}}+\\
& + H\left( {s,t,u} \right){\delta _{ad}}{\delta _{bc}} \,. \this\label{GeneralAplitude}
\end{align*}
This amplitude can be decomposed in an Isospin invariant basis as
\begin{align}
{M^{ab,cd}} = \sum\limits_{m = 0}^2 {{T^m}P_m^{abcd}} \,, \label{IsospinAmpDef}
\end{align}
where $T^m$ are the Isospin amplitudes and the basis vectors \cite{Bjorken1965} are
\begin{align}
P_0^{abcd}& = \frac{1}{3}{\delta _{ab}}{\delta _{cd}} \,,\\ P_1^{abcd} &= \frac{1}{2}\left( {{\delta _{ac}}{\delta _{bd}} - {\delta _{ad}}{\delta _{bc}}} \right) \,,\\
P_2^{abcd}& = \frac{1}{2}\left( {{\delta _{ac}}{\delta _{bd}} + {\delta _{ad}}{\delta _{bc}}} \right) - \frac{1}{3}{\delta _{ab}}{\delta _{cd}} \,.
\end{align}
Expanding the amplitude, Eq.\eqref{IsospinAmpDef},
and identifying coefficients of the Kronecker delta between Eq.\eqref{GeneralAplitude} and Eq.\eqref{IsospinAmpDef}, leads to a system of linear equations with the solution
\begin{align}
{T^0} &= 3 \,F\left( {s,t,u} \right) + G\left( {s,t,u} \right) + H\left( {s,t,u} \right) \,,\\
{T_1} &= G\left( {s,t,u} \right) - H\left( {s,t,u} \right) \,,\\
{T_2} &= G\left( {s,t,u} \right) + H\left( {s,t,u} \right) \,.
\end{align}
The Isospin amplitudes with rho-meson exchange are given by
\begin{align*}
T_A^0 =4i{e^2}{m^2} \biggl[& \frac{{2 + \left( {3 + z} \right)R}} {{2{m^2}\left( {1 - z} \right)R + {M^2}}}+ \\ &+\frac{{2 + \left( {3 - z} \right)R}}{{2{m^2}\left( {1 + z} \right)R + {M^2}}} \biggr] \,,\this
\end{align*}
\begin{align*}
T_A^2 
 &=  - \frac{1}{2}T_A^0 \, \this
\end{align*}
\begin{align*}
T_A^1   &= 2i{e^2}{m^2}\Biggl[ \frac{{4zR}}{{4{m^2}\left( {1 + R} \right) - {M^2}}} + \frac{{2 + \left( {3 + z} \right)R}}{{2{m^2}\left( {1 - z} \right)R + {M^2}}} \\
  &- \frac{{2 + \left( {3 - z} \right)R}}{{2{m^2}\left( {1 + z} \right)R + {M^2}}} \Biggr] \,, \this
\end{align*}
The Isospin amplitudes with Higgs-exchange become
\begin{align*}
T_H^0 =  16i{\kappa ^2}{\nu ^2} &\Biggl[ \frac{1}{{2{m^2}\left( {1 - z} \right)R + m_H^2}} -\frac{3}{{4{m^2}\left( {1 + R} \right) - m_H^2}} \\
  &+ \frac{1}{{2{m^2}\left( {1 + z} \right)R + m_H^2}} \Biggr] \,,\this \\
T_H^1= 16i{\kappa ^2}{\nu ^2} &\biggl[ \frac{1}{{2{m^2}\left( {1 - z} \right)R + m_H^2}} +\\
  &- \frac{1}{{2{m^2}\left( {1 + z} \right)R + m_H^2}} \biggr] \,,\this \\
T_H^2 = 16i{\kappa ^2}{\nu ^2} &\biggl[ \frac{1}{{2{m^2}\left( {1 - z} \right)R + m_H^2}} +\\
  &+ \frac{1}{{2{m^2}\left( {1 + z} \right)R + m_H^2}} \biggr] \,.\this
\end{align*}
The Isospin amplitudes corresponding to the four-pion vertex are given by
\begin{align}
T_\lambda ^0 &=  - 5\, i \,{\lambda _4} & T_\lambda ^1 & =0 &T_\lambda ^2 & =  - 2 \,i\, {\lambda _4} \,.
\end{align}
The scattering lengths are computed from the coefficients of the partial wave scattering amplitude obtained from the projection of the Isospin amplitudes over the Legendre polynomials. The Isospin amplitude 
${T^I}\left( {{q^2},z} \right)$ expressed as a sum over partial wave scattering amplitudes $T_m^I\left( q^2 \right)$ is
\begin{align}
{T^I}\left( {{q^2},z} \right) = 32\,\pi \,\sum\limits_{m = 0}^\infty  {\left( {2m + 1} \right){P_m}\left( z \right)T_{m}^I\left( {{q^2}} \right)} \,,
\end{align} 
where ${P_m}\left( z \right)$ are the Legendre polynomials, with $z = \cos \theta$. From the orthogonality of the Legendre polynomials, 
the partial-wave scattering amplitudes can be extracted by projecting ${T^I}\left( {{q^2},z} \right)$ onto the Legendre polynomials
\begin{align*}
T_n^I\left( {{q^2}} \right) &= \frac{1}{{64\pi }}\int\limits_{ - 1}^1 {{P_n}\left( z \right){T^I}\left( {{q^2},z} \right)dz} \,. \this
\end{align*}
These amplitudes can be expanded in a power series in terms of $q^2$ \cite{Colangelo2001125} as 
\begin{align}
T_n^I 
&= i \,R^n \left[ {a_n^I + b_n^I R +  \ldots } \right] \,,
\end{align}
where $R=\dfrac{q^2}{m^2}$ is defined in Eq.\eqref{zcostheta}, and $a_n^I$ and $b_n^I$ are the scattering lengths given by
\begin{align}
a_0^0 &= \frac{{{e^2}}}{{2\pi }}\frac{{{m^2}}}{{{M^2}}} + \frac{{{\kappa ^2}{\nu ^2}}}{\pi }\left[ {\frac{1}{{m_H^2}} - \frac{3}{{2\left( {4{m^2} - m_H^2} \right)}}} \right]+\nonumber\\ &- \frac{{5{\lambda _4}}}{{32\pi }} \\
b_0^0 &= \frac{{{e^2}}}{{4\pi }}\frac{{{m^2}}}{{{M^2}}}\left( {3 - 4\frac{{{m^2}}}{{{M^2}}}} \right)+ \nonumber\\
&+ \frac{{2{m^2}{\kappa ^2}{\nu ^2}}}{\pi m_H^4 }\left[ {\frac{3m_H^4}{{{{\left( {4{m^2} - m_H^2} \right)}^2}}} - 1} \right]\\
a_1^0 &= 0 \qquad b_1^0 = 0\\ 
a_2^0 &= \frac{{2{e^2}}}{{15\pi }}\frac{{{m^4}}}{{{M^4}}}\left( {1 + 4\frac{{{m^2}}}{{{M^2}}}} \right) + \frac{{16}}{{15\pi }}\frac{{{m^4}{\kappa ^2}{\nu ^2}}}{{m_H^6}} \\ 
b_2^0 &= \frac{{4{e^2}}}{{5\pi }}\frac{{{m^6}}}{{{M^6}}}\left( {1 - 12\frac{{{m^2}}}{{{M^2}}}} \right) - \frac{{96}}{{5\pi }}\frac{{{m^8}{\kappa ^2}{\nu ^2}}}{{m_H^8}}\\ \nonumber\\ 
a_0^1 &= 0 \qquad b_0^1 = 0\\ 
a_1^1 &= \frac{{{e^2}}}{{24\pi }}\frac{{16{m^6} - 3{m^2}{M^4}}}{{4{m^2}{M^4} - {M^6}}} + \frac{2}{{3\pi }}\frac{{{m^2}{\kappa ^2}{\nu ^2}}}{{m_H^4}} \\
b_1^1 &= \frac{{{e^2}}}{{6\pi }}\dfrac{{{m^4}\left( { 1- \dfrac{64}{3}\dfrac{m^6}{M^6} + 16\dfrac{m^4}{M^4} - 4\dfrac{m^2}{M^2}} \right)}}{{{M^4}{{\left( 1- 4\dfrac{m^2}{M^2} \right)}^2}}}+\nonumber\\ &- \frac{8}{{3\pi }}\frac{{{m^4}{\kappa ^2}{\nu ^2}}}{{m_H^6}}\\
a_2^1 &= 0 \qquad b_2^1 = 0\\
\nonumber\\
a_0^2 &=  - \frac{{{e^2}}}{{8\pi }}\frac{{{m^2}}}{{{M^2}}} + \frac{{{\kappa ^2}{\nu ^2}}}{{2\pi m_H^2}} - \frac{{{\lambda _4}}}{{32\pi }} \\ b_0^2 &=  - \frac{{{e^2}}}{{48\pi }}\frac{{{m^2}}}{{{M^2}}}\left( {3 - 4\frac{{{m^2}}}{{{M^2}}}} \right) - \frac{1}{{3\pi }}\frac{{{m^2}{\kappa ^2}{\nu ^2}}}{{m_H^4}}\\
a_1^2 &= 0 \qquad b_1^2 = 0\\
a_2^2 &=  - \frac{{{e^2}}}{{30\pi }}\frac{{{m^4}}}{{{M^4}}}\left( {1 + 4\frac{{{m^2}}}{{{M^2}}}} \right) + \frac{8}{{15\pi }}\frac{{{m^4}{\kappa ^2}{\nu ^2}}}{{m_H^6}} \\ b_2^2 &=  - \frac{{{e^2}}}{{15\pi }}\frac{{{m^6}}}{{{M^6}}}\left( {1 - 12\frac{{{m^2}}}{{{M^2}}}} \right) - \frac{{16}}{{5\pi }}\frac{{{m^6}{\kappa ^2}{\nu ^2}}}{{m_H^8}} \,.
\end{align}
The experimental input \cite{Patrignani} for the average masses of the charged and neutral pions and rho-mesons is 
$m =0.1372734 \pm  0.0000007 ~\rm{GeV}$, and $M =0.77649 \pm 0.00034~\rm{GeV}$. The $\rho\pi\pi$ coupling, $e \equiv g_{\rho\pi\pi}$ in standard VMD \cite{SAKURAI1960} equals the rho-meson leptonic decay constant \cite{Patrignani}, $f_\rho = 4.97 \pm 0.07$, as determined from its leptonic decay rate.  However, data on the electromagnetic form factor of the pion \cite{BESIII}, as well as theory (Large $N_c$ - QCD) \cite{CAD} shows a substantial deviation $e = (1.21 \pm 0.02) f_\rho$. Taking this into account gives $e = 6.0 \pm 0.1$.
The value of the vacuum expectation value $\nu$ can be inferred from the definition of the mass of the rho-meson, becoming
\begin{align*}
\nu =0.130 \pm 0.004 ~\rm{GeV} \,.\this
\end{align*}
The pion decay constant is \cite{Patrignani} $F_\pi = 92.1 \pm 1.2\rm{MeV}$. The four-pion coupling is  \cite{Scherer2003}
\begin{equation}
\lambda_4  = \left(\frac{m}{F}\right)^2 = 2.45074\pm 0.1568 \,.\this
\end{equation}
The mass of the symmetry breaking field $m_H$ is taken to be the mass of the $f_0(500)$-meson \cite{Pelaez2016}
\begin{align}
m_H & = m_{\rm{f_0}}=0.450 \pm 0.016 ~\rm{GeV} \,.
\end{align}
This choice is motivated by the fact that the energy range of validity of this non-Abelian theory is of order ${\mathcal{O}}(1 \, 
{\mbox{GeV}})$, rather than the mass scale of the actual Higgs boson. Using the above values of the parameters and those for $a^0_0$ and $b^0_0$, one extracts an averaged value for $\kappa$ 
\begin{align}
\kappa = 1.31 \pm 0.03 \,.
\end{align}
The results for the scattering lengths are shown in Table I, together with some predictions from Chiral Perturbation Theory (CHPT), and available experimental values.

\begin{widetext}

\begin{table}[H]
	\begin{center}
		\begin{tabular}{@{}cccccccc@{}}
			
			\toprule
Scattering &Weinberg&CHPT (LO)&CHPT (NLO) &This work&CHPT&Experiment\\
Lengths & \cite{Weinberg1966} &\cite{donoghue2014dynamics} &\cite{donoghue2014dynamics} & &\cite{Colangelo2001125},\cite{Bijnens2000} &\cite{donoghue2014dynamics},\cite{data_sl} \\
			\midrule
			\midrule
$a^0_0$ &$0.20$  	&$0.16$	 &$0.20$				&$0.21$ 					&$0.220$				 &$0.220\pm0.005$ \\
$b^0_0$ &			&$0.18$	 &$0.26$				&$0.30$ 					&$0.276$				 &$0.25\pm0.03$\\


$a^0_2\times 10^{3}$ & 			&$0$&$2$&$2.06$		& $1.75$	  & $1.7\pm 3$ \\
$b^0_2\times 10^{4}$ &		    &   &   & $-5.23$ 	& $-3.55$   \\
			
			\midrule 
			

$a^1_1$ &			  	&$0.030$ &$0.036$				&$0.0528$				&$0.0379$	 &$0.038 \pm 0.002 $\\
$b^1_1$ &				&$0$	 &$0.043$ 				&$0.0053$				&$0.0057$	\\
	
			
			\midrule
			
$a^2_0$ & $-0.06$			 &$-0.045$&$-0.041$		&$-0.0456$				&$-0.0444$		& $-0.044\pm 0.001$ \\
$b^2_0$ &			 &  	  &  			&$-0.0225$				&$-0.0803$		&$-0.082\pm0.008$\\


$a^2_2\times 10^{4}$ & 			  	&$0$ 	&$3.5$ 		& $-2.03$	& $1.70$ 			&$1.3\pm 3$ \\

			\bottomrule
		\end{tabular}
	\end{center}
	\medskip
		\raggedright
			\caption{Summary of predicted values of the scattering lengths at tree-level in this theory, together with other determinations from CHPT, and available experimental data \cite{donoghue2014dynamics}, \cite{data_sl}.}
	\label{treeresultstablelisted}
\end{table}
\end{widetext}

\section{Discussion and Conclusion}
The Abelian KLZ theory of pionic interactions \cite{KLZ1967}, at leading order in perturbation theory, was successfully applied to determine the electromagnetic pion form factor in the time-like region (from the rho-meson self energy) \cite{GALE1991}. This form factor agreed well with experimental data, as it coincided with the Gounaris-Sakurai expression \cite{GS} in the vicinity of the rho-peak.
Further applications of this theory are the pion electromagnetic form factor in the space-like region \cite{CADWillers2007}, from the one-loop triangle diagram, the scalar radius of the pion \cite{CADWillers2008}, and the scalar form factor of the pion in the space-like region \cite{CADLushozi2015}. 
Regarding the one-loop triangle diagram determining the Abelian KLZ electromagnetic pion form factor \cite{CADWillers2007}, it should be noticed that it involves an order $\cal{O}$$ (g^2)$ correction to the tree-level diagram. In spite of the strength of the $\rho\pi\pi$ coupling, this correction is mild due to the dimensional regularization overall factor $1/(4 \, \pi)^2$. The agreement of the pion form factor with data in the wide range $- q^2 = 0.01 - 10.0 \, {\mbox{GeV}^2}$ is excellent, as witnessed by a chi-squared per degree of freedom $\chi^2 = 1.1$.\\
On the basis of this success, and given the need for a strong interaction theory of pionic interactions, covering the energy region between CHPT (around threshold) and below the onset of QCD, we proposed here a non-Abelian extension of the KLZ theory. This new theory has the added advantage of being renormalizable. Mass generation was achieved by invoking the Higgs mechanism, with a Higgs boson mass comparable to the mass of the $f_0$(500)-meson, i.e. in the region of applicability of the theory.
To test its reliability, pion-pion scattering lengths were determined at tree-level, in reasonable agreement with data, and various CHPT predictions. This provides encouraging support for this strong interaction theory of pionic interactions. Further applications would involve the next order in perturbation theory which, however, is beyond the scope of this work.\\

\section{Acknowledgements}
This work was supported in part by the National Research Foundation (South Africa), the National Institute for Theoretical Physics (South Africa), the University of cape Town (South Africa), the Alexander von Humboldt Foundation (Germany), and the Deutsche Forschungsgemeinschaft (Germany). Discussions with Hubert Spiesberger are greatly appreciated.

\end{document}